\documentclass[twocolumn,aps,prb,10pt,showpacs,superscriptaddress]{revtex4}
\usepackage{graphics}
\usepackage{graphicx}
\usepackage{amsmath}
\usepackage{amssymb}
\usepackage{subfigure}
\usepackage{longtable}
\usepackage{multirow}
\usepackage{bm}
\usepackage{hyperref}

\newcommand{\ve}[1]{\bm{\mathrm{#1}}}

\newcommand{\w}{\omega}

\begin{document}
\title{Conventional and Acoustic Surface Plasmons on Noble Metal Surfaces: \\ A Time-dependent Density Functional Theory Study}

\author{Jun Yan}
\email{junyan@stanford.edu}
\affiliation{SUNCAT Center for Interface Science and Catalysis,  
SLAC National Accelerator Laboratory \\ 2575 Sand Hill Road, Menlo Park, CA 94025, USA}
\affiliation{Center for Atomic-scale Materials Design, Department of
Physics \\ Technical University of Denmark, DK - 2800 Kgs. Lyngby, Denmark}
\author{Karsten W. Jacobsen}
\affiliation{Center for Atomic-scale Materials Design, Department of
Physics \\ Technical University of Denmark, DK - 2800 Kgs. Lyngby, Denmark}
\author{Kristian S. Thygesen}
\affiliation{Center for Atomic-scale Materials Design, Department of
Physics \\ Technical University of Denmark, DK - 2800 Kgs. Lyngby,
Denmark}
\affiliation{Center for Nanostructured Graphene (CNG), Department of
Physics \\ Technical University of Denmark, DK - 2800 Kgs. Lyngby,
Denmark}
\date{\today}

\begin{abstract}
First-principles calculations of the conventional and acoustic surface plasmons (CSPs and ASPs) on the (111) surfaces of Cu, Ag, and Au are presented. The effect of $s-d$ interband transitions on both types of plasmons is investigated by comparing results from the local density approximation and an orbital dependent exchange-correlation (xc) potential that improves the position and width of the $d$ bands. The plasmon dispersions calculated with the latter xc-potential agree well with electron energy loss spectroscopy (EELS) experiments. For both the CSP and ASP, the same trend of  Cu$<$Au$<$Ag is found for the plasmon energies and is attributed to the reduced screening by interband transitions from Cu, to Au and Ag. This trend for the ASP, however, contradicts a previous model prediction. While the ASP is seen as a weak feature in the EELS, it can be clearly identified 
in the static and dynamic dielectric band structure.  
\end{abstract}

\maketitle

The collective electronic excitations at surfaces, known as surface plasmons, have generated extraordinary interest in the past decade due to the diverse potential applications in sensing, imaging, surface enhanced spectroscopy, catalysis and solar energy harvesting\cite{Plasmonics}.  
Noble metals such as gold and silver are widely used in experimental plasmonics due to their stability and well controlled plasmonic properties. Besides the widely investigated 'conventional' surface plasmons (CSPs), which have finite plasmon energies for all wave vectors\cite{Liebsch_book}, the noble metal surfaces support another type of surface excitations, namely the acoustic surface plasmons (ASPs)\cite{Echenique_B04, Echenique_B05}. The ASP exhibits a linear dispersion of plasmon energy ($\omega_{\text{asp}}$) as a function of wave vector with $\omega_{\text{asp}} \rightarrow 0$ as $\ve q \rightarrow 0$. It was predicted to exist on surfaces supporting Shockley surface states within the bulk electronic band gap and in a two dimensional electron gas on top of a substrate\cite{Echenique_B04, Echenique_B05}. Experimentally, the ASP has been identified on Be(0001)\cite{Be0001_Nature}, Cu(111)\cite{Pohl_EPL10} and  Au(111)\cite{Palmer_L10} surfaces, as well as in graphene adsorbed on substrates\cite{Seyller_B08, Cupolillo_Carbon12}. 

Theoretically, the CSP and ASP of noble metal surfaces have been treated using two different models. The CSP was modeled using a jellium surface, taking the screening by $d$ electrons into account via an effective dielectric constant\cite{Liebsch_L93}. For the ASP, a one-dimensional (1D) model potential was constructed to reproduce the main features of the Shockley surface states (SS) and their underlying bulk states\cite{Echenique_B05}. However, these model calculations do not account for the presence of $d$ states and are not fully {\it ab-initio}. 

While experiments are so advanced that not only the plasmons on the clean noble metal surfaces can be measured\cite{Palmer_L10, Pohl_EPL10}, but their behavior in the presence of adsorbates\cite{Palmer_L09} and disorder\cite{Carminati_L10} has also be studied, a unified {\it ab-initio} treatment of both kinds of plasmons is yet missing.

In this paper, we use time-dependent density functional theory with the adiabatic local density approximation (ALDA) to calculate the CSP and ASP of the (111) surfaces of the noble metals Cu, Au and Ag. Single-particle energies and orbitals are obtained using the GLLBSC xc-potential to get quantitatively correct $d$ band positions and plasmon energies\cite{Jun_AgH}. The calculated dispersions for both the CSP and ASP are in good agreement with available experimental EELS measurements on Cu and Au surface, despite the fact that the calculated ASP energies for Au(111) are underestimated by more than 0.5 eV for $\mathrm{q}>0.2$~\AA$^{-1}$. While the CSPs are strongly influenced by the $s$-$d$ interband transitions, the ASPs are only slightly redshifted. For the ASP energies, we found a trend of Cu$<$Au$<$Ag, contradicting the trend derived from previous model calculations Ag$<$Cu$<$Au\cite{Echenique_B05}. Finally, we demonstrate that the ASP can be clearly identified from the static and dynamic dielectric band structure of the noble metal surfaces.

All calculations were performed using the projector-augmented wave method \textsc{gpaw}\cite{GPAW_10, Jun_response} code. The (111) surfaces were modeled by slabs of either 10 or 24 atoms thickness (ca. 2 or 5 nm), for the study of CSP and ASP, respectively. The thicker slabs are essential to converge the ASP in the long wave length limit where interaction between the plasmons on the two sides of the slab becomes more significant. For comparison, the ground state calculations were performed using both the LDA and GLLBSC\cite{GLLB, GLLBSC} functionals. The plasmon energies were obtained as the peaks in the loss function 
\begin{equation}S(\ve q, \omega) = -\mathrm{Im\epsilon^{-1}_{\mathbf{G}=0,\mathbf{G}^{\prime}=0}(\mathbf{q}, \omega)},
\end{equation}
where $\ve G$ and $\ve G'$ are reciprocal lattice vectors and  $\epsilon^{-1}_{\mathbf{G}\mathbf{G}^{\prime}}(\mathbf{q}, \omega)$ is the inverse dielectric matrix. The latter is obtained using linear response time-dependent density functional theory\cite{Jun_response} in the adiabatic local density approximation and with single-particle energies and orbitals from either the LDA or GLLBSC ground state calculations. More details on the calculations can be found in Ref. \footnote{Experimental lattice parameters of Au (4.08~\AA),  Ag (4.09~\AA) and Cu (3.61~\AA) were adopted.  The supercell used contains 30 \AA~vacuum and is discretized using uniform mesh with grid spacing 0.18 \AA.  The structure relaxation was carried out using LDA and $8 \times 8$ Monkhorst-Pack {\it k}-point sampling.   To construct the density response function, a dense k-point sampling of $100 \times 100$ and $80 \times 80$ in the surface Brillouin zone and unoccupied bands of 30 and 20 eV above Fermi level are used in calculations for obtaining the CSP and ASP, respectively.  Two kinds of local field corrections are employed in the response function calculations: an isotropic 150 eV for all three directions and an anisotropic one with 10 eV within the surface plane and 500 eV perpendicular to the surface. Both give essential the same plasmon dispersions. Based on the above parameters, the energies of the CSP and ASP achieve convergence of 0.05 eV. }

\begin{figure}[t]
  \centering
  \includegraphics[width=0.95\linewidth,angle=0]{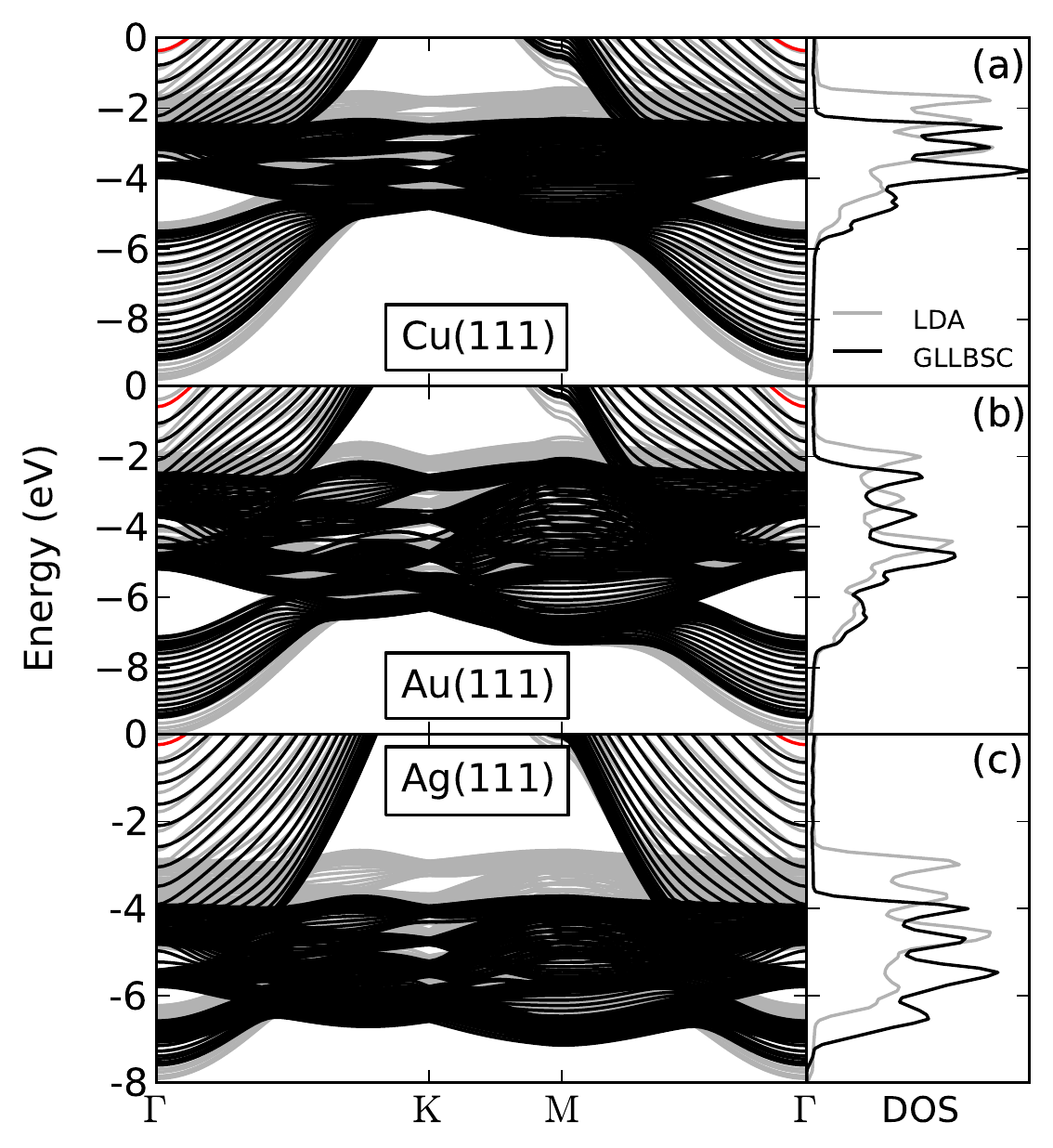}
  \caption[]{ \label{Fig:Bandstr} (Color online) Band structure (left panel) and density of states (DOS, right panel) for the (111) surfaces, represented by 24 layer slabs ($\sim$ 5nm thickness), of (a) Cu, (b) Au and (c) Ag, calculated with LDA (grey lines) and GLLBSC (black lines).  The bands marked in red represent Shockley surface states (SS).  Zero energy indicates the Fermi level.  }
\end{figure}

Figure \ref{Fig:Bandstr} shows the band structure (left panel) of the 24 layer slabs used to represent the (111) surfaces of Cu, Ag and Au. The corresponding density of states are presented in the right panel. For each surface the GLLBSC xc-potential lowers the upper edge of the $d$ bands and leads to a narrowing of the $d$ band width, compared to LDA. The lowering and narrowing of the $d$ bands with respect to the LDA calculations were also observed in previous GW calculations on bulk Cu\cite{Cu_GW} and Ag\cite{Ag_GW}.  The calculated GLLBSC upper/lower edge of the $d$ bands are at around -2/-6 eV (Cu), -2/-8 eV (Au) and -4/-7 eV (Ag). These values are in good agreement with angle resolved photoemission (ARPES) measurement on polycrystalline Cu and Au surfaces\cite{ARPES_noblemetal} and Ag(111) films \cite{Ag_PES}. The broadening of the $d$ bands from Cu(111) to Au(111) (with an increase of the $d$ band width from 4 to 6 eV), and the downshifting of the $d$ bands from Au(111) to Ag(111) (with the upper edge lowers from -2 to -4 eV) indicate reduced $d$ electrons screening from Cu, to Au and Ag surface.

 \begin{figure*}[t]
  \centering
  \includegraphics[width=1.0\linewidth,angle=0]{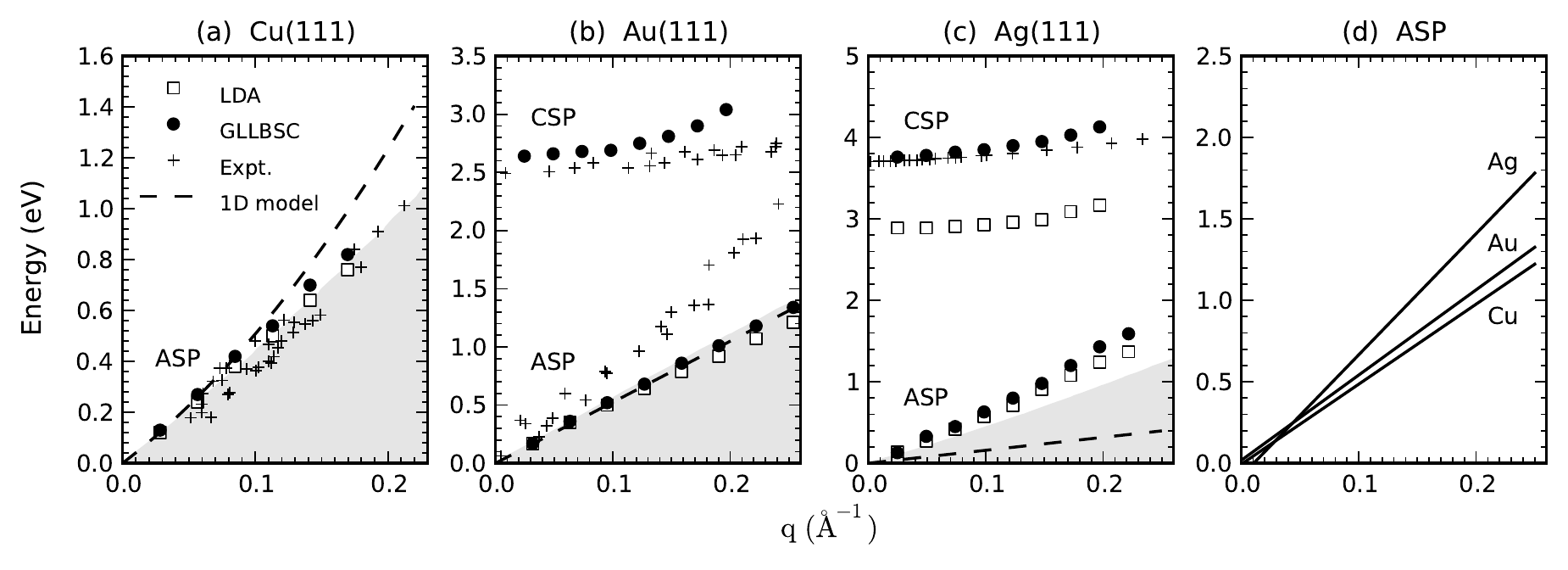}
  \caption[]{ \label{Fig:Dispersion} Conventional and acoustic surface plasmon (CSP and ASP) dispersions for the (111) surfaces of Cu, Au, and Ag calculated with LDA (hollow squares) and GLLBSC (solid dots). The plasmon energies are obtained as peaks in the loss function $S(\ve q, \omega)$, the examples of which are shown in Fig. \ref{Fig:dielectric_w} as black lines. Previous model results (dashed line) and EELS experimental data (crosses)\cite{Pohl_EPL10} are shown for comparison. Shaded areas mark the Shockley electron-hole pair continuum calculated with GLLBSC.  (d) Summary of the ASP dispersion obtained with the GLLBSC potential for each of the three surfaces.   }
\end{figure*}

Figure \ref{Fig:Dispersion} (a)-(c) show the CSP and ASP dispersions for the (111) surfaces of Cu, Au and Ag, respectively. The results of the present work are indicated by squares (LDA) and circles (GLLBSC). The CSP is not visible on the Cu(111) surface [panel (a)] due to strong screening (Landau damping) of the plasmon by the $s-d$ interband transitions. On Au(111) [panel (b)] the CSP appears just above 2.5 eV in GLLBSC calculations, but is absent in the LDA calculated loss spectrum. The fact that no CSP is predicted by LDA is due to the overestimation of the $d$ band position which leads to overscreening of the plasmons\cite{Jun_AgH}. On Ag(111) [panel (c)] the CSP appears for $q=0$ at around 3 eV and 4 eV for LDA and GLLBSC, respectively, and disperses towards higher energy for larger $q$. Overall, this shows a trend of Cu(nonexistent)$<$Au$<$Ag for the CSP energies. Comparing to the experimental results (crosses), 
the CSP energies obtained with GLLBSC presents a significant improvement over the LDA. This is clearly a consequence of the improved description of the $d$ band positions with the GLLBSC.
In contrast, the ASP dispersions shown in Fig. \ref{Fig:Dispersion} are much less influenced by the $d$ bands, showing less than 0.1 eV difference between the LDA and GLLBSC results for all surfaces.  The ASP energies  exhibit the same trend of Cu$<$Au$<$Ag [see panel (d)], as found for the CSP. 

In general, the $s-d$ interband transitions can have two effects on the surface plasmons depending on the relative energies of the two types of excitations. If the energies of the $s-d$ transitions are comparable to that of the plasmon, the latter is strongly redshifted and damped. In contrast, if the plasmon energy is significantly lower than the $s-d$ transitions, only a slight redshift of the plasmon energy is observed\cite{Jun_B08}. The first scenario clearly applies to the CSP while the second describes the case for the ASP. While the trend of Cu(nonexistent)$<$Au$<$Ag observed for the CSP can be explained by the same trend in the $d$ band positions (broadening and downshifting), it is not obvious why the ASP energies exhibit the same trend. 

Before analyzing further the observed trend in the ASP energies, we compare our {\it ab-initio} ASP dispersions with the model predictions of Ref. \onlinecite{Echenique_B05} and EELS experiments. For Cu(111) in Fig. \ref{Fig:Dispersion} (a), the GLLBSC presents a slight improvement over the model by predicting somewhat lower ASP energies, although in this case both the model and LDA are already in quite good agreement with experiments\cite{Pohl_EPL10}. These ASP energies lie just above the upper boundary of the electron-hole (e-h) pair continuum, the latter agrees with previous calculations\cite{Crampin_B05}. 
For Au(111), the GLLBSC result for the ASP agrees well with experiments at small momentum transfer, but deviates at larger $\mathbf{q}$ exhibiting a too small slope for the plasmon dispersion. Similar to that for Cu(111), the calculated ASP dispersion for Au(111) coincides with the upper boundary of the Shockley e-h pair continuum. Since Au(111) undergoes a surface reconstruction with a $22\times \sqrt{3}$ unit cell\cite{Palmer_L10}, we investigated the effect of strain by compressing the Au lattice constant by 4.4\%. Insignificant  changes (around 0.1 eV blueshift) were found for the ASP energies. A possible reason for the discrepancy could be the spin-orbit coupling, which splits the surface state bands into two subbands in $\ve k$-space\cite{Bruno_JPCM04}, but is out of scope of this work.  For Ag(111), the GLLBSC ASP dispersion has a significantly larger slope than the model prediction and is around 0.2-0.4 eV (corresponds to $\mathrm{q}=0.1$-$0.2$~\AA$^{-1}$) above the e-h pair continuum. A previous {\it ab-initio} calculations on the ASP of a Be(0001) surface also shows that the calculated ASP lies slightly above the e-h pair continnum\cite{Be0001_Nature}.  To the best of our knowledge, no experimental results are available for the ASP on Ag(111). Compared to Au(111), the Ag(111) surface does not undergo a surface reconstruction (although easily oxidized) and spin-orbit coupling effects are negligible. Consequently, experimental results for Ag(111) would be highly important to advance our understanding of the surface plasmons on noble metal surfaces.

The trend of Cu$<$Au$<$Ag predicted for the ASP energies contradicts the model prediction of Ag$<$Cu$<$Au. 
In the model, the ASP energies are directly correlated with the Fermi velocity of the Shockley surface band. 
By rigidly shifting the surface bands in energy relative to the Fermi level (by up to 0.4 eV) in our {\it ab-initio} calculations we are able to keep the effective mass of the surface bands unchanged while altering the binding energy and thus the Fermi velocity of the surface states. However, we found that the ASP energies have very weak dependence on the binding energy and the Fermi velocity of the surface states. On the other hand, according to Fig. \ref{Fig:Dispersion}, the ASPs seem to be highly correlated with the upper edge of the e-h pair continnum. However, the upper edge of the e-h pair continuum differ by at most 0.1 eV (at $\mathrm{q}=0.2$~\AA$^{-1}$) among the three metal surfaces and thus is not sufficient to account for the calculated difference in the ASP energies at the same $\mathrm{q}$. Furthermore, the discrepancy, shown in Fig. \ref{Fig:Dispersion} (c), between the {\it ab-initio} and the model calculations for the e-h pair continuum of the Ag surface, suggests that 
a 1D harmonic approximation for the surface bands might not be sufficient in describing the e-h pair continuum in the case of Ag. 
 In order to understand the trend of the ASP energies, we introduce in the following the static and dynamic dielectric band structure of the surfaces.

 \begin{figure}[b]
  \centering
  \includegraphics[width=1.0\linewidth,angle=0]{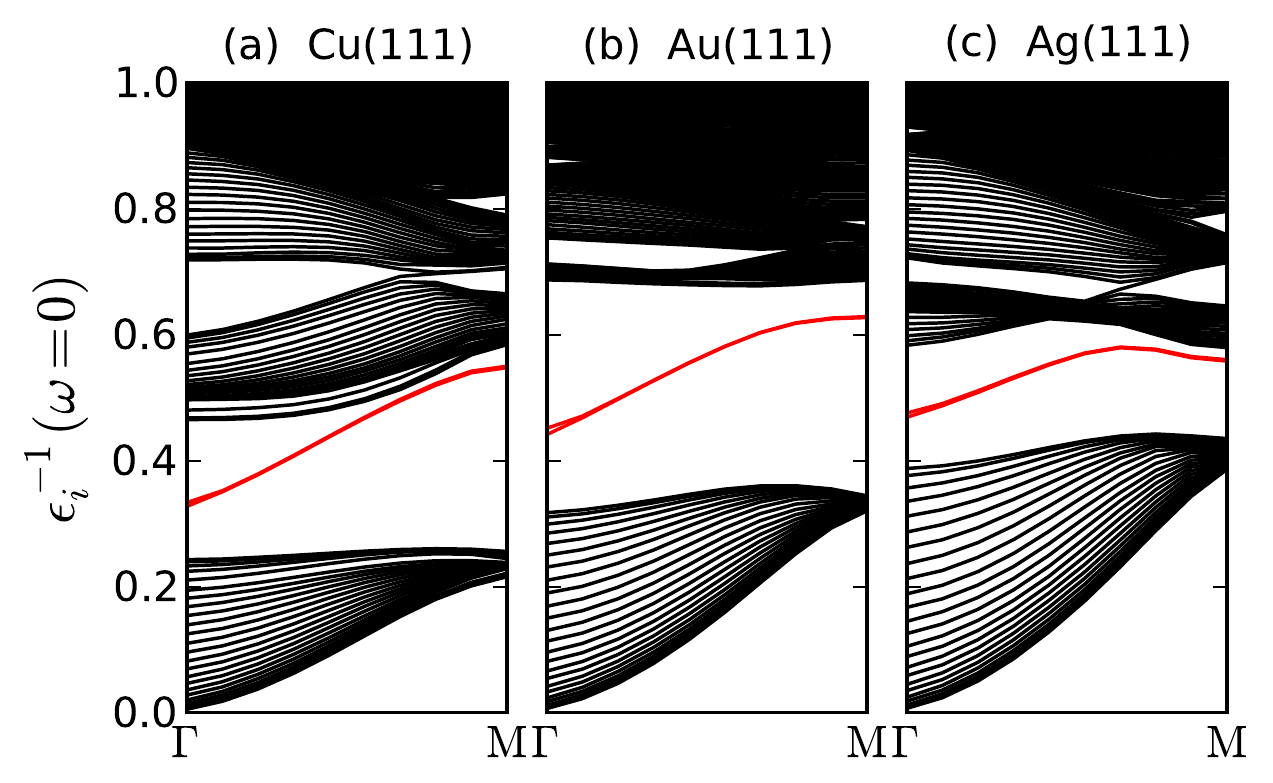}
  \caption[]{ \label{Fig:dielectric_band} (Color online) Static dielectric band structure of the (111) surfaces of Cu, Au, and Ag represented by 24 layer slabs ($\sim$ 5nm thickness). $\epsilon_i^{-1}$ are the real eigenvalues of the static inverse microscopic dielectric matrix $\epsilon^{-1}_{\mathbf{G} \mathbf{G}^{\prime}}$. The bands marked in red represent the surface modes (SMs). }
\end{figure}

Fig. \ref{Fig:dielectric_band} (a)-(c) show the static 
 dielectric band structure $\epsilon_i^{-1}(\mathbf{q})$ for the 24 layer slabs representing the (111) surfaces of Cu, Au and Ag, respectively. Following Ref. \onlinecite{Tosatti_79}, $\epsilon_i^{-1}(\mathbf q)$ are obtained as eigenvalues by diagonalizing its inverse dielectric matrix $\epsilon^{-1}_{\mathbf{G} \mathbf{G}^{\prime}} (\mathbf{q}, \omega=0)$ at each $\mathbf{q}$ in the irreducible Brillouin zone (BZ) .
The dielectric band structures of all three metal slabs consist of two almost continuous groups of bands separated by a gap from around 0.2-0.5 (Cu), 0.3-0.7 (Au) and 0.4-0.6 (Ag). The density of the individual bands lying within these groups increases for thicker slabs and  are consequently related to the bulk. In contrast, the single band marked by red lying inside the gap of each surface does not change as the slab is made thicker. It implies that this band relates to the surface state of the slab. To verify, we checked the corresponding eigenvector of the red band and found it localized at the surface of the slab decaying exponentially into the slab, which is the characteristic of a surface state. In fact, the red band is rather two degenerate bands in agreement with the fact that the slab contains two identical surfaces. Furthermore, we have calculated the dielectric band structure of an Al(111) 20 layer slab which does not support electronic surface states. The dielectric band structure (not shown) presents only one group of dense bands, i.e. there is no gap, and no bands that can be related to the surface. Therefore, the existence of the red band (from hereon denoted the surface mode (SM)) is a consequence of the SS in the electronic band structure. While the electronic SS (red lines in Fig. \ref{Fig:Bandstr}) exist only at small $\ve q$ and merge into the bulk bands at larger $\ve q$, the SMs (red lines in Fig. \ref{Fig:dielectric_band}) are much more prominent and are visible inside the bulk dielectric gap up to the zone boundary. 

 \begin{figure}[t]
  \centering
  \includegraphics[width=1.0\linewidth,angle=0]{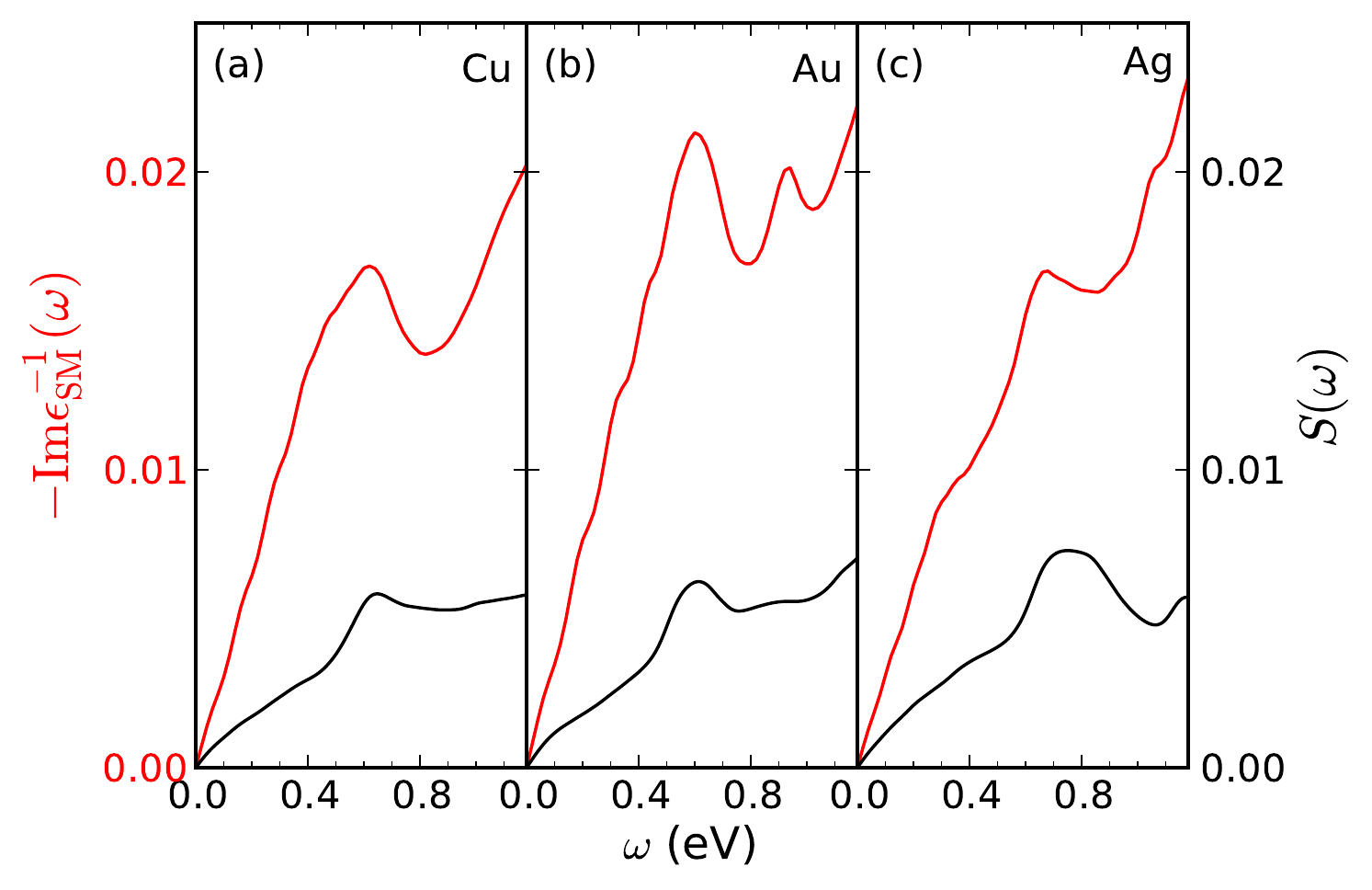}
  \caption[]{ \label{Fig:dielectric_w} (Color online) Imaginary part of the dynamic dielectric eigenvalues $-\mathrm{Im}\epsilon_i^{-1}(\mathbf{q}, \omega$) for the surface states ($i = \mathrm{SM}$, red lines in Fig. \ref{Fig:dielectric_band}) plotted as a function of $\omega$ at $|\mathbf{q}|=0.125/$\AA~for (a) Cu(111), (b) Au(111) and (c) Ag(111), respectively. They are compared to the macroscopic loss function $S(\mathbf{q}, \omega$) (black lines) at the same $\ve q$. 
  }
\end{figure}

We have argued above that the SM in the dielectric band structure is a consequence of the SS in the electronic band structure. In the following we present further proof of this statement.  Figure \ref{Fig:dielectric_w} shows the imaginary part of the dynamic dielectric eigenvalues $-\mathrm{Im}\epsilon_i^{-1}(\ve q, \w)$ for the surface mode ($i = \mathrm{SM}$) at a particular $|\ve q|=0.125/$\AA. The dynamic dielectric eigenvalues are calculated for each $\omega$ following the same diagonalizing procedure as for the static case. The macroscopic loss functions $S(\mathbf{q}, \omega)$, which can be directly compared to the EELS spectra,  are plotted as well. For each surface, $-\mathrm{Im}\epsilon_{\mathrm{SM}}^{-1}(\ve q, \w)$ resembles the loss function and exhibits a peak at the ASP energy. Such a peak in $-\mathrm{Im}\epsilon_{\mathrm{i}}^{-1}(\ve q, \w)$ is absent for all the other bands ($i \neq \mathrm{SM}$) in the dynamic dielectric band structure. This suggests that the ASP is indeed originating from the SM. 

Finally, we propose an explanation for the trend of Cu$<$Au$<$Ag observed for the ASP energies. Shown in Fig. \ref{Fig:dielectric_band}, for each $\ve q$, the value of $\epsilon_i^{-1}$ corresponding to the SM exhibits a trend of Cu$<$Au$<$Ag. Recalling that $\epsilon_i^{-1}$ gives the total potential due to an externally applied potential, this shows that the electronic screening for the SM (thus ASP) is strongest on Cu(111), weaker on Au(111) and weakest on Ag(111). Again, this is a consequence of the position and width of the $d$ bands as discussed in connection with the CSP.
Furthermore, the existence of the SM within the bulk dielectric gap up to the BZ boundary agrees with ASP being observed at relatively large $\ve q$.

In conclusion, the conventional and acoustic surface plasmons on the (111) surfaces of Cu, Au and Ag were investigated using time-dependent density functional theory with the adiabatic local density approximation. Our {\it ab-initio} results agree well with available EELS experiments. The energies of both conventional and acoustic surface plasmons follow the same trend of Cu $<$ Au $<$Ag. The trend is attributed the reduced $d$ bands screening from Cu, to Au and Ag. Experiments of the ASP energies on Ag(111) surface will be very important to test our findings and further advance our understanding of plasmons and electronic screening at metal surfaces. 

J. Yan acknowledges support by the Department of Energy, Office of Basic Energy Sciences, under contract DE-AC02-76SF00515. The Center for Nanostructured Graphene is sponsored by the Danish National Research Foundation. The Catalysis for Sustainable Energy initiative is
funded by the Danish Ministry of Science, Technology and Innovation. 
The computational studies were supported as part of the 
Center on Nanostructuring for Efficient Energy Conversion, an Energy 
Frontier Research Center funded by the U.S.
Department of Energy, Office of Science, Office of Basic Energy 
Sciences under Award No. DE-SC0001060.


\bibliographystyle{apsrev}

\end{document}